\documentclass{ifacconf}

\usepackage{graphicx}      
\usepackage{natbib}        

\newtheorem{mydef}{Definition}

\newtheorem{myprob}{Problem}

\newtheorem{mypro}{Proposition}

\newtheorem{remark}{Remark}

\usepackage{amsfonts}
\DeclareMathSymbol{\shortminus}{\mathbin}{AMSa}{"39}
\usepackage{enumitem}
\usepackage{multirow}
\usepackage{makecell}
\usepackage{booktabs}
\usepackage{array}
\usepackage{graphicx}
\usepackage{subfigure} 
\usepackage{float}
\usepackage{amsmath}
\usepackage{amsfonts}
\usepackage{epsfig}
\usepackage{graphicx}
\usepackage{arydshln}
\usepackage{algpseudocode}
\usepackage{verbatim}
\usepackage{subfigure}
\usepackage{enumerate}
\usepackage{rotating}
\usepackage{color}
\usepackage{flushend}
\usepackage{mathrsfs}
\usepackage{amssymb}
\usepackage[ruled,vlined,linesnumbered]{algorithm2e}
\usepackage{float}
\usepackage{soul}
\usepackage{fancyhdr}
\usepackage{url}
\usepackage{makecell}
\begin{document}
\begin{frontmatter}

\title{Model Predictive Control \\ for  Signal Temporal Logic Specifications \\ with Time Interval Decomposition\thanksref{footnoteinfo}} 

\thanks[footnoteinfo]{This work was supported by the National Key Research and Development Program of China (2018AAA0101700) and the National Natural Science Foundation of China (62061136004, 61803259, 61833012).}

\author[1]{Xinyi Yu} 
\author[1]{Chuwei Wang} 
\author[1]{Dingran Yuan} 
\author[1]{Shaoyuan Li}
\author[1]{Xiang Yin} 

\address[1]{Department of Automation, Shanghai Jiao Tong University, Shanghai 200240, China. (E-mail: \{yuxinyi-12, wangchuwei, geniustintin, syli, yinxiang\}@sjtu.edu.cn)
}

\begin{abstract}     
   In this paper, we investigate the problem of Model Predictive Control (MPC) of dynamic systems for high-level specifications described by Signal Temporal Logic (STL) formulae. Recent works show that MPC has the great potential in handling logical tasks in reactive environments. However, existing approaches suffer from the heavy computational burden, especially for tasks with large  horizons. In this work, we propose a computationally more efficient MPC framework for STL tasks based on time interval decomposition. Specifically, we still use the standard shrink horizon MPC framework with Mixed Integer Linear Programming (MILP) techniques for open-loop optimization problems. 
   However, instead of applying MPC directly for the entire task horizon, 
   we decompose the STL formula into several sub-formulae with disjoint time horizons, and shrinking horizon MPC is applied for each short-horizon sub-formula iteratively. To guarantee the satisfaction of the entire STL formula and to ensure the recursive feasibility of the iterative process,  we introduce new terminal constraints to connect each sub-formula.  We show how these terminal constraints can be computed by an effective inner-approximation approach. The computational efficiency of our approach is illustrated by a case study. 
\end{abstract}

\begin{keyword}
   Signal Temporal Logic; Model Predictive Control; Feasible Sets.
\end{keyword}

\end{frontmatter}

\section{Introduction} 
    Decision-making under dynamic environment is one of the central problems in control of Cyber-Physical Systems (CPS). In the past years, there has been growing interest in controller synthesis for    high-level complex tasks.  Specifically, temporal logics, such as Linear Temporal Logic (LTL) and Signal Temporal Logic (STL), provide   expressive and user-friendly tools for formal description and automated design of complex tasks involving both continuous variables and discrete logics.

	Signal temporal logic was firstly developed in \cite{maler2004monitoring} for the purpose of behavior monitoring.
	STL formulae are evaluated over continuous time signals. Compared with LTL, the semantics of STL are quantitative, and therefore, provide a measure for  the degree of the satisfaction or violation in addition to the Boolean satisfaction. 
	Recently, STL has been successfully applied to the analysis and control of many engineering CPS including, e.g., autonomous robots   \cite{lindemann2017robust}, intelligent transportation systems \cite{mehr2017stochastic}  and smart buildings \cite{ma2020sastl}.
	
	In the context of control synthesis for STL specifications, one of the most widely used approaches is to encode the satisfaction of STL formulae as mixed-integer constraints  \cite{raman2014model}. Then the STL control problem can be solved by applying Mixed Integer Linear Programming (MILP) techniques together with Model Predictive Control (MPC) framework. The encoding-based approach has also been adopted in  \cite{Alexandre2010, sadraddini2015robust, farahani2017shrinking}. The main advantage of this approach is that the solution is complete and globally optimal. However, since one needs to introduce decision variables for each time instant, the encoding-based approach  suffers from the heavy computational burden, especially for tasks with large horizons.
	
	To mitigate the high  complexity in STL control synthesis problem, several computationally efficient methods have been developed recently in the literature. The techniques include, e.g., gradient-based optimizations \cite{mehdipour2019arithmetic,gilpin2020smooth,leung2020back}, control barrier functions  \cite{lindemann2018control}, 
	prescribed performance control  \cite{lindemann2017prescribed, varnai2019prescribed},  referenced way-points \cite{sun2022multi} and computationally tractable robustness \cite{lindemann2019robust}. Furthermore,  learning-based methods have also been developed for ensuring STL specifications, e.g., \cite{li2019formal,liu2021recurrent}. However, compared with the encoding-based MPC approach, the above mentioned methods are usually not complete and cannot provide formal guarantees for the existence of a solution when facing complex formula.
	In the context of encoding-based methods,  \cite{kurtz2022mixed} proposed a more efficient encoding approach where disjunction can be encoded using a logarithmic number of binary variables and conjunction can be encoded without binary variables.
	In \cite{sadraddini2018formal}, the authors showed that binary variables can be further reduced by writing formulae in positive normal forms.
	However, these improved encoding methods still suffer from the huge computational complexity when the task horizon increases.

    In this paper,  we also focus on STL control synthesis using shrinking horizon MPC together with variable encoding  techniques.  Compared with existing MPC approach that applies directly to the entire task horizon, the contributions of this paper are as follows:
	\begin{itemize}[leftmargin=*] 
		\item
		We decompose the STL formula into several sub-formulae with disjoint time horizons. Then shrinking horizon MPC is applied for each sub-formula iteratively with short-horizons, which significantly reduces computational complexity of the entire synthesis process. 
		\medskip
		\item
		Note that, by focusing on each sub-formula only, the subsequent task may become infeasible.
		In order to ensure the recursive feasibility of the entire process, we introduce terminal constraint sets between  connected sub-formulae. This guarantees the satisfaction of the global STL task. 
		Also, we provided an effective method for offline computation of the inner-approximation the terminal constraint sets. 
		The  computational complexity does not increase exponentially as the horizon increases.\medskip
		\item
		Finally, we present a case study of robot motion planning to demonstrate the efficiency of the proposed framework. 
		We show that, compared with the full horizon MPC approach,  our approach is more scalable for STL formula with  long task horizon. 
	\end{itemize}


	The rest of the paper is organized as follows.
	We provide some necessary preliminaries in Section~\ref{sec:pre} and formulate the problem  in Section~\ref{sec:MPC}. In section~\ref{sec:sshmpc}, we present our main algorithm   and detailed computations of feasible sets are presented in Section~\ref{sec:feasibleset}. A case study is presented in Section~\ref{sec:case} to show the scalability of our approach, Finally,  we conclude the paper in Section~\ref{sec:conclu}.

\emph{Notations: }We denote by $\mathbb{R}$ and $\mathbb{Z}_{\geq 0}$ the sets of the real numbers and non-negative integers, respectively. We use non-bold letters to denote scalars or column vectors and use bold letters to denote sequences. 
The Minkowski sum of sets $A$ and $B$ is  $A \oplus B = \{a+b \mid a \in A, b\in B\}$.

\section{Preliminary}\label{sec:pre}

	\subsection{System Model}
	We consider a discrete-time control system of form
	\begin{equation}\label{eq:system}
		x_{k+1} = f(x_k, u_k) + w_k,
	\end{equation}
	where 
	$x_k \!\in\! \mathcal{X}\subseteq \mathbb{R}^n$ is the state at instant $k$, 
	$u_k \!\in\! \mathcal{U}\subseteq \mathbb{R}^m$ is the control input at instant $k$, 
	$w_k \!\in\! \mathcal{W}\subseteq \mathbb{R}^n$ is the external input or disturbance at instant $k$
	and 
	$f:\mathcal{X}\times \mathcal{U} \times \mathcal{W}\to \mathcal{X}$ is a function describing the dynamic of the system. 
	We assume that the disturbances are unknown but  belong to a given compact set $\mathcal{W}$. 
	Also, we assume that the dynamic function $f$ is Lipschitz continuous on $x$, i.e., for all $x, x' \!\in\! \mathcal{X}, u \!\in\! \mathcal{U}$, there exists positive constant $L$ such that 
	$ |f(x', u) - f(x, u)|  \leq L|x' - x|$.

	Suppose that the system is in state $x_k \!\in\!  \mathcal{X}$ at instant $k \!\in\! \mathbb{Z}_{\geq 0}$. 
	Then given a sequence of control inputs
	$\mathbf{u}_{k:T-1}=  u_k u_{k+1} \cdots u_{T-1}  \!\in\! \mathcal{U}^{T-k}$
	and 
	a sequence of disturbances
	$\mathbf{w}_{k:T-1}=  w_k w_{k+1} \cdots w_{T-1}  \!\in\! \mathcal{W}^{T-k}$, 
	the  solution of the system is a sequence of states
	$\xi_f(x_k,\mathbf{u}_{k:T-1},\mathbf{w}_{k:T-1} ) = x_{k+1} \cdots x_{T} \!\in\! \mathcal{X}^{T-k}$ such that $x_{i+1}=f(x_i,u_i)+w_i,\forall i=k,\cdots, T-1$, and the solution of nominal system is denoted by $\xi_f(x_k,\mathbf{u}_{k:T-1})$ similarly.
	Also, we denote by $\xi_f^T(x_k,\mathbf{u}_{k:T-1},\mathbf{w}_{k:T-1} )=x_T$ the last state in the sequence. 
	
	\subsection{Signal Temporal Logic}
	Given a finite sequence of states $\mathbf{x}$, we use signal temporal logic (STL) formulae \cite{maler2004monitoring} with bounded-time temporal operators to describe whether  or not the trajectory of the system satisfies some high-level properties. 
	Formally, the syntax of STL formulae is as follows 
	\[
	\Phi ::=   \top\mid \pi^\mu \mid \neg \Phi \mid \Phi_1 \wedge \Phi_2 \mid \Phi_1 \textbf{U}_{[a,b]} \Phi_2,
	\]
	where $\top$ is the \textsf{true} predicate, $\pi^\mu$ is a predicate whose truth value is determined by the sign of its underlying predicate function $\mu:\mathbb{R}^n \to \mathbb{R}$  and it is true if $\mu(x_k) \geq 0$; otherwise it is false.
	Notations $\neg$ and $\wedge$ are the standard Boolean operators ``negation" and ``conjunction", respectively,  which can further induce ``disjunction" by $\Phi_1 \vee \Phi_2:=\neg(\neg \Phi_1 \wedge \neg \Phi_2)$
	and ``implication" by $\Phi_1 \to \Phi_2:= \neg \Phi_1 \vee   \Phi_2$. 
	Also, $\textbf{U}_{[a,b]}$ is the temporal operator ``\emph{until}", where $a,b\in \mathbb{R}$. 
	
	STL formulae are evaluated on state sequences.  
	We use notation $(\mathbf{x},k) \models \Phi$ to denote that sequence $\mathbf{x}$ satisfies STL formula $\Phi$ at instant $k$. 
	The reader is referred to \cite{maler2004monitoring} for more details on the semantics of STL formulae. 
	Particularly, we have 
	$(\mathbf{x},k)\models \pi^\mu$ iff $\mu(x_k)\geq 0$ 
	and 
	$(\mathbf{x},k)\models \Phi_1 \mathbf{U}_{[a,b]} \Phi_2$ 
	iff $\exists k' \!\in\! [k+a, k+b]$ such that $(\mathbf{x},k') \models \Phi_2$
	and  $\forall k'' \!\in\! [k, k']$, we have $(\mathbf{x},k'') \models \Phi_1$.
	Furthermore, 
	we can also induce temporal operators  
	\begin{itemize}
		\item 
		``\emph{eventually}" $\textbf{F}_{[a,b]} \Phi:= \top \mathbf{U}_{[a,b]} \Phi$
		such that it holds when $(\mathbf{x},k) \models \Phi$ for some $k'\in [k+a,k+b]$; and\medskip
		\item 
		``\emph{always}" $\textbf{G}_{[a,b]} \Phi:=\neg \textbf{F}_{[a,b]} \neg \Phi$ such that it holds  when $(\mathbf{x},k) \models \Phi$ for any $k'\in [k+a,k+b]$.
	\end{itemize}

	We write $\mathbf{x} \models \Phi$ whenever $(\mathbf{x}, 0) \models \Phi$.

	Given an STL formula $\Phi$,   it is well-known that the satisfaction of  $\Phi$ can be completely determined only by those states within its \emph{horizon}.  
	Hereafter, we use notation $\Phi^{[S_\Phi,T_\Phi]}$ to emphasize that formula $\Phi$ only depends on time horizon  $[S_\Phi,T_\Phi]$, 
	where $T_\Phi$ is the terminal instant of $\Phi$ which is the maximum sum of all nested upper bounds 
	and   $S_\Phi$ is the starting instant of $\Phi$ which is the minimum time instant that appears in the formula.
	For example, for $\Phi = \textbf{F}_{[2,7]}x^{\mu_1} \wedge  \textbf{G}_{[3,12]} x^{\mu_2}$, we have $T_\Phi = \max\{7,12\} = 12$ and $S_\Phi = \min\{2, 3\} = 2$. 
	 With a slight abuse of notation, hereafter, we also use a broader interval as its superscript, e.g., $\Phi^{[S'_\Phi, T'_\Phi]}$ with $S'_\Phi \leq S_\Phi$ and $T'_\Phi \geq T_\Phi$.

	In addition to the Boolean satisfaction, we can also evaluate an STL formula \emph{quantitatively} according to the robust semantics. 
	Formally, for any STL formula $\Phi$, state sequence $\mathbf{x}$ and time instant $k$, we denote by $\rho^\Phi_{\mathbf{x},k}$ the \emph{space-robustness function}  the same as \cite{Alexandre2010}:
	
	In this paper, we consider the following restrict but expressive enough fragment of STL formulae:
	\begin{subequations} \label{eq:stl}
		\begin{align}
			\varphi ::= \top \mid \pi^\mu \mid \neg \varphi \mid \varphi_1 \wedge \varphi_2, \qquad \qquad  \\
			\Phi::= \textbf{F}_{[a,b]} \varphi \mid \textbf{G}_{[a,b]} \varphi \mid \varphi_1\textbf{U}_{[a,b]} \varphi_2 \mid \Phi_1 \wedge \Phi_2,
		\end{align}
	\end{subequations}
	where $\varphi_1, \varphi_2$ are formulae of class $\varphi$, and $\Phi_1, \Phi_2$ are formulae of class $\Phi$.  
	Specifically, we only allow the temporal operators to be applied once for Boolean formula and nested temporal operators are not allowed. 

	\subsection{Feasible Set of Signal Temporal Logic}
	In our previous works \cite{yu2022online, yu2022model}, we propose a method to compute the so called \emph{feasible set} of STL formula for a dynamic system without considering disturbances. 
	Here we  briefly review this concept and interested readers are referred to   \cite{yu2022model} for details.

	We observe that formula $\Phi$  of form $\eqref{eq:stl}$ can be written as 
	\begin{equation}\label{eq:stl-seg}
		\Phi = \bigwedge_{i=1}^{N} \Phi_i^{[a_i, b_i]}, \vspace{-6pt}
	\end{equation}
	where 
	$N$   denotes the total number of sub-formulae, 
	and for
	each sub-formula $\Phi_i^{[a_i, b_i]}$, it is effective within  time interval $[a_i, b_i]$ and is in the form of 
	$\mathbf{G}_{[a_i, b_i]} x \!\in\! \mathcal{H}_i$, $\mathbf{F}_{[a_i, b_i]} x \!\in\! \mathcal{H}_i$ or $x \!\in\! \mathcal{H}_{i}^1 \mathbf{U}^{\prime}_{[a_i, b_i]} x\! \in\! \mathcal{H}_{i}^2$. 
	We denote by $\mathcal{I} = \{1, \cdots, N\}$ the index set of all sub-formula and by $O_i \!\in\! \{\mathbf{G}, \mathbf{F}, \mathbf{U}'\}$ the unique temporal operator in $\Phi_i$.
	Also, we denote by $\mathcal{I}_k=\{i \!\in\! \mathcal{I} \mid a_i \leq k \leq b_i\}$ the index set of sub-formulae that are effective at instant $k$. Similarly, we denote by $\mathcal{I}_{<k}=\{i \!\in\! \mathcal{I} \mid b_i <k\}$ and $\mathcal{I}_{>k}=\{i \!\in\! \mathcal{I} \mid k <a_i \}$ the index sets of sub-formulae that are effective strictly before and after instant $k$ respectively.
	
	Let $I\subseteq \mathcal{I}$ be a set of indices representing those sub-formulae that have not yet been satisfied. 
	We say  $I$ is a remaining set at  instant $k$ if 
	(i)   $\mathcal{I}_{<k}\cap I=\emptyset$; and 
	(ii)  $\mathcal{I}_{>k}\subseteq I$; and 
	(iii) $\{ i \!\in\! \mathcal{I}_k \mid [O_i=\mathbf{G}] \vee [O_i=\mathbf{U}' \wedge k=a_i]\} \subseteq I$.  
	We denote by $\mathbb{I}_k$ the set of all possible remaining set at instant $k$. 
	
	Then  given a remaining set $I$ at instant $k$, we call  formula 
	\begin{equation}
	    \hat{\Phi}_{k}^I =\bigwedge_{i \in I\cap \mathcal{I}_k} \Phi_i^{[k,b_i]} \wedge \bigwedge_{i \in \mathcal{I}_{>k}} \Phi_i^{[a_i,b_i]} \nonumber
	\end{equation}
	the $I$-remaining formula representing the entire task remained. 
	The set of states from which the remaining task can be fulfilled is called the $I$-remaining feasible sets.

	\begin{mydef}[$I$-Remaining Feasible Sets]\label{def:feasibleset}\upshape
		Given an STL formula $\Phi$ of form~\eqref{eq:stl-seg}, 
		a subset of indices $I\subseteq \mathcal{I}$ at time instant $k$, 
		the $I$-remaining feasible set at instant $k$, denoted by 
		$X_k^{I} \subseteq \mathcal{X}$, is the set of states from which there exists a solution that satisfies the $I$-remaining sub-formula at $k$, i.e.,
		\begin{align}\label{eq:feasibleset}
			X_k^{I} \! =\! 
			\left\{ 
			x_k \in \mathcal{X} \,\middle\vert\, \!\!\!\!
			\begin{array}{cc}
				\exists \ \mathbf{u}_{k:T_\Phi-1} \in \mathcal{U}^{T_\Phi-k} \\
				\text{ s.t. } x_k \xi_f(x_{k}, \mathbf{u}_{k:T_\Phi-1}) \models \hat{\Phi}_{k}^I
			\end{array} 
			\right\}.\vspace{-6pt}
		\end{align}
	\end{mydef}

	Suppose that $I$ is a remaining set at instant $k$, 
	we denote by $\textsf{succ}(I,k)=\{I'\subset I\mid I'\in \mathbb{I}_{k+1}\}$ the set of all possible \emph{successor sets} of $I$. Intuitively, a transition from $I$ to  $I'$ means that sub-formulae in $I\setminus I'$ have been satisfied currently. 
	This means that the system should be in the following region at instant $k$ 
	\begin{equation}\label{eq:H}
		H_k(I, I') = \bigcap_{i \in I \cap \mathcal{I}_k} H_i, \nonumber
	\end{equation}
	with	
	\begin{align}\label{eq:three-case}
		H_i = 
		\left\{
		\begin{array}{cl}
			\mathcal{H}_i^1 \cap \mathcal{H}_i^2 \ \ \ 
			& \text{if}  \ i \!\in\! \textsf{sat}_\textsf{U}(I,I')\\
			\mathcal{H}_i^1   \setminus \mathcal{H}_i^2  
			& \text{if} \  
			O_i\!=\!\mathbf{U}' \wedge 
			i \!\notin\!   \textsf{sat}_\textsf{U}(I,I')  \\
			\mathcal{H}_i
			& \text{if} \   O_i\!=\!\mathbf{G},
		\end{array}
		\right.  \nonumber 
	\end{align} 
	where 
	\begin{equation}
		\textsf{sat}_\textsf{U}(I,I')
		=
		\{
		i\in I:  O_i=\mathbf{U}'\wedge i\notin I' 
		\}.  \nonumber
	\end{equation}

	Then  the $I$-remaining feasible set can be computed 	by the following results \cite{yu2022model}.
	
	\begin{thm}\label{thm:feasible set}
		$I$-remaining feasible set $X_k^I$ defined in Definition~\ref{def:feasibleset} for the time instant $k$ can be computed as follows
		\begin{equation}\label{eq:dynpro}
			X_k^I = \bigcup_{I' \in \textsf{succ}(I, k)} \Big( H_k(I, I') \cap \Upsilon(X_{k+1}^{I'}) \Big), 
		\end{equation}
	where $\Upsilon(\cdot)$ is the one-step set defined by: for any $\mathcal{S}\subseteq \mathcal{X}$
	\begin{equation}\label{eq:onestep}
		\Upsilon(\mathcal{S}) = \{x \in \mathcal{X} \mid \exists u \in \mathcal{U}  \text{ s.t. }  f(x, u) \in \mathcal{S}\}.  \nonumber
	\end{equation}
	\end{thm}

\section{Model Prediction Control for Signal Temporal Logic}\label{sec:MPC}
	
Our objective is to synthesize a feedback control strategy  such that the sequence generated by the closed-loop system satisfies the desired STL formula under all possible disturbances. Furthermore, we want to minimize control effort while maximizing the control performance. 
	
Formally, given  state $x_k$ at instant $k$ and input sequence $\mathbf{u}_{k:T-1}=u_k u_{k+1}\cdots u_{T-1}$, 
	we consider a generic cost function 
	\[
	J: \mathcal{X}\times \mathcal{U}^{T-k}\to \mathbb{R}.
	\]
	For example, $J$ can be defined as the nominal satisfaction degree of the STL formula without disturbance, i.e., 
	\[
		J(x_k, \mathbf{u}_{k:T-1})
		=  -\rho^\Phi_{  \mathbf{x}_{0:k} \xi_f(x_k,\mathbf{u}_{k:T-1}),k},
	\]
	or it can be defined as the worst-case satisfaction degree of the STL formula for all possible disturbances, i.e., 
	\[
		J(x_k, \mathbf{u}_{k:T-1})
		=\underset{\mathbf{w}_{k:T-1}\in  \mathcal{W}^{T-k}}{\max}  -\rho^\Phi_{\mathbf{x}_{0:T}^w,k},
	\]
	where $\mathbf{x}_{0:T}^w=\mathbf{x}_{0:k}\xi_f(x_k,\mathbf{u}_{k:T-1},\mathbf{w}_{k:T-1} )$, or it can be the total energy of the control inputs, i.e.,
	\[
		J(x_k, \mathbf{u}_{k:T-1})
		= \Sigma_{i=k}^{T-1} u_i^2.
	\]
	
	Our approach for solving the STL control synthesis problem follows the basic framework of model predictive control. 
	Specifically, at each instant, we solve a finite-horizon optimization problem to compute a finite open-loop sequence of inputs such that 
	the cost function is minimized subject to the constraints on both the system's dynamic and the STL formula. Note that we only apply the first input in the computed sequence to the system. Then  at the next instant,  we recompute the input sequence based on the actually measured current-state and repeat until the terminal instant. 
	Formally, the optimization problem for each instant is formulated as follows. 
	
	\begin{myprob}[\textbf{Robust STL Optimization Problem}]\label{prob1}\upshape
		Given system in   the form of  Equation~(\ref{eq:system}), an STL formula $\Phi$, a cost function $J$, 
		the current state $x_k\in \mathcal{X}$ at instant $k$, 
		previous state sequence $\mathbf{x}_{0:k-1}$
		and some constant $T$, 
		find an optimal input sequence  $\mathbf{u}^*_{k:T-1}$ that minimizes the cost function subject to  constraints on the system's dynamic and the temporal logic requirement. 
		Formally, we have the following optimization problem
		\begin{subequations}\label{second:main} 
			\begin{align}
				& \underset{\mathbf{u}_{k:T-1}}{\text{minimize}}  & & J(x_k, \mathbf{u}_{k:T-1} )\\
				& \text{subject to}\nonumber
				& & \\
				& &&
				\!\!\!\!\!\!\!\!\!\!\!\!\!\!\!\!\!\!\!\!\!\!\!\!\!\!\!\!\!\!\!\!
				\forall \mathbf{w}_{k:T-1}\!\in\! \mathcal{W}^{T \shortminus k}: \mathbf{x}_{0:k} \xi_f(x_k,\mathbf{u}_{k:T \shortminus 1},\mathbf{w}_{k:T \shortminus 1})\models \Phi, \\
				& &&
				\!\!\!\!\!\!\!\!\!\!\!\!\!\!\!\!\!\!\!\!\!\!\!\!\!\!\!\!\!\!\!\!
				u_k,u_{k+1},\cdots, u_{T-1}\in \mathcal{U} 
			\end{align} 
		\end{subequations} 
	\end{myprob}
	
	\begin{remark}\upshape\label{rmk:methods}
		Effective approaches have been proposed in the literature for solving the above optimization problem. 
		To handle the STL satisfaction constraint, a basic approach is proposed in \cite{raman2014model} by encoding state sequence as well as the satisfaction of the formula using binary variables. 
		Then the logical-constrained optimization problem is converted to a Mixed Integer Linear Program (MILP). 
		For the purpose of robust satisfaction under disturbances, based on the MILP-based approach in \cite{raman2014model},    \cite{raman2015reactive} further purposes a counterexample-guided inductive synthesis  (CEGIS) scheme that finds a robust optimal solution iteratively.  
		In \cite{farahani2015robust}, more efficient approaches for solve the robust optimization problem have been provided. 
		Furthermore, for discrete-time linear system, \cite{sadraddini2015robust} proposes an efficient framework to synthesize control strategies using positive normal forms of STL. In this work, we will adopt the CEGIS-based approach in  \cite{raman2015reactive} for solving Problem~1 and its variant. 
		Note that other approaches aforementioned can also be adopted in principle. 
	\end{remark} 
	
In the MPC framework,  only the first control input $u_k^*$ in the optimal input sequence $\mathbf{u}_{k:T\shortminus 1}^*$ computed will be applied to the system. 
Depending on the actual disturbance $w_k$ occurs, the controller will measure the new state $x_{k+1}$ at the next instant and recompute the optimal sequence based on \texttt{Problem~1}. 
This procedure is formally provided in \texttt{Algorithm~1}, which is referred to as the \emph{shrinking horizon MPC} because the optimization horizon of \texttt{Problem~1} is shrinking from $T_\Phi$ to $1$ as time instant $k$ increases. 
Here, we assume that Problem~\ref{prob1} is feasible at the initial instant $k=0$ for the initial state $x_0$; otherwise,  the set of feasible initial states can be calculated by the method that will be introduced in Section~\ref{sec:feasibleset}. 

\IncMargin{1em}
\begin{algorithm}[ht]
	\caption{Shrinking Horizon MPC for STL}
	\KwIn{STL formula $\Phi$,  dynamic system model of form ($\ref{eq:system}$)   and cost function $J$}
	\KwOut{Control input $u_k$ at each instant $k$.\vspace{3pt}}
	$T \gets T_{\Phi}$ and $k \gets 0$ \\
	\While{$k < T$}
	{
		measure current state $x_k$  \\ 
		solve \texttt{Problem~1} based on $\mathbf{x}_{0:k}$, $T$ and $\Phi$ and obtain optimal inputs $\mathbf{u}_{k:T\shortminus 1}^*= u_k^*u_{k+1}^*  \cdots u_{T \shortminus 1}^*$ \\
		apply control input $u_k^*$ \\
		$k \gets k+1$
	}
\end{algorithm}

	The shrinking horizon MPC provides an effective framework for solving the reactive control synthesis problem for both STL specification and performance optimization. 
	However, the complexity for solving \texttt{Problem~1} at each instant grow exponentially as the prediction horizon increases.
	Therefore, the main computational challenging for shrinking horizon MPC is in the first few instants since we need to predict almost the entire horizon the formula, 
	which may make this approach computationally infeasible for STL tasks with large horizons.

	\section{Time Interval Decomposition Framework}\label{sec:sshmpc}

	\subsection{Time Interval Decomposition}
	As we discussed above, the main computational challenging in shrinking horizon MPC is the predication horizon of the task. 
	In this paper, we still use shrinking horizon MPC as the basis.   
	However, we propose a new  time interval decomposition framework, 
	which leverages  structural properties of STL formulae in terms of time interval decomposition to further reduce the computational complexity. 
	Specifically, we assume that the STL formula of interest can be further divided into a set of disjoint time intervals. 
	Then we solve the shrinking horizon MPC problem for each sub-formula with smaller time horizon.  Finally, by putting each time intervals together ``correctly", we solve the entire STL control synthesis problem. 

	To motivate our approach, let us consider the following STL formula	
	\[
	\Phi = (\textbf{G}_{[0,100]} \pi^{\mu_1} ) \wedge (\textbf{F}_{[21,50]}  \pi^{\mu_2} ) \wedge (\textbf{F}_{[81,100]} \pi^{\mu_3} ).
	\]
	For the above formula, the time horizon is $T_\Phi=100$. Therefore, using the standard shrinking horizon MPC, we need to compute an optimization problem with predication horizon $100$ initially, which is computationally very challenging. 
	However, we also note that the above formula can be written equivalently as a conjunction of four sub-formulae with disjoint time intervals
	\[
	\Phi  = \Phi_1^{[0,20]} \wedge \Phi_2^{[21,50]} \wedge \Phi_3^{[51,80]} \wedge \Phi_4^{[81,100]},
	\]
	where $\Phi_1 = \textbf{G}_{[0,20]} \pi^{\mu_1}$, $\Phi_2 = (\textbf{G}_{[21,50]} \pi^{\mu_1}) \wedge (\textbf{F}_{[21,50]} \pi^{\mu_2})$, $\Phi_3 = \textbf{G}_{[51,80]} \pi^{\mu_1}$ and $\Phi_4 = (\textbf{G}_{[81,100]} \pi^{\mu_1}) \wedge (\textbf{F}_{[81,100]} \pi^{\mu_3})$. 
	As an alternative, we can enforce the satisfaction of the entire STL formula $\Phi$ by enforcing the satisfaction of each sub-formula $\Phi_i$. 
	For example, initially, we enforce $\Phi_1$ using shrinking horizon MPC by setting the state terminal instant to $T_{\Phi_1}=20$ (control input terminal instant is 19). 
	Then at instant $k=20$, we start to enforce $\Phi_2$ using shrinking horizon MPC by setting the state terminal instant to $T_{\Phi_2}=50$, and so forth. 
	Therefore, we  solve four MPC problems with prediction horizons at most $30$ instants,
	which is much more easier than solving a single  MPC problem with predication horizon at most $100$. 
	
	To formalize the above motivation, we assume that the STL formula considered can be written as the conjunction of a set of sub-formulae whose effective horizons are disjoint, 
	i.e., $\Phi$ is of the following form
	\begin{equation}\label{eq:separation}
		\Phi^{[S_\Phi,T_\Phi]} = \Phi_1^{[S_{1},T_{1}]} \wedge \Phi_2^{[S_{2},T_{2}]}  \wedge \cdots \wedge \Phi_N^{[S_{N},T_{N}]}
	\end{equation}
	where $S_{1}=S_\Phi, T_{N}=T_\Phi$ and for each $i=1,\cdots,N-1$, we have $T_{i} \leq S_{i+1}$.
	
	\begin{remark}\upshape
	The above assumption of time interval decomposition is without loss of generality since we can always take $N=1$. But no computational reduction will be gained for this case.  However, in many practical examples, different requirements in the global task are natural time disjoint. For this scenario, one can decompose the formulae into more than one sub-formulae. 
	\end{remark}

	\subsection{Robust Optimizations with Terminal Constraints}
	As we mentioned above, at each instant $k$, suppose that $k \!\in\! [T_{i-1},T_{i}-1]$,
	our approach is to enforce the $i$th task $\Phi_i$ by solving the robust optimization problem with predication horizon from $k$ to $T_i$ as \texttt{Problem~1}. 
	Since each sub-formula $\Phi_i$ is satisfied within its effective horizon  and the entire formula is their conjunction, then the global task $\Phi$ is satisfied. 
	
	However, the main issue of this approach is that how we can guarantee the \emph{recursive feasibility} during the entire control process. 
	More specifically, for $k \!\in\! [T_{i-1},T_{i}-1]$, we consider the optimization problem for $\Phi_i$ only up to $T_i$. 
	This, however, may lead to an optimal solution ending up with a state from which some subsequent formula $\Phi_j,j>i$ cannot be satisfied. 
	Therefore, for each sub-task period, in addition to the constraints in  \texttt{Problem~1}, we   also need to take the feasibility of all subsequent formulae into consideration. 
	
	To this end, we define $\mathcal{T}_i$ as the set of states from which the subsequent task $\bigwedge_{i<j\leq N} \Phi_j$ is feasible in the sense that no matter what the disturbance sequence is, there exists at least one control input sequence such that all the subsequent STL formulae can be satisfied. 
	Formally, for $i=1,\cdots,N-1$, we define
	\begin{align}\label{eq:terminal}
		&\mathcal{T}_{i} =  \\
		&\left\{
		x_{T_{i}} \!\in\! \mathcal{X} \,\middle\vert\,
		\begin{array}{cc}
				\exists   \mathbf{u}_{T_{i}:T_{\Phi}\shortminus 1} \!\in\! \mathcal{U}^{T_{\Phi}\shortminus T_{i}},
				\forall \mathbf{w}_{T_{i}:T_{\Phi}\shortminus 1} \!\in\! \mathcal{W}^{T_{\Phi} \shortminus T_{i}},   \\
			\text{s.t. }  	\xi_f(x_{T_{i}}, \mathbf{u}_{T_{i}:T_{\Phi} \shortminus 1}, \mathbf{w}_{T_{i}:T_{\Phi}\shortminus 1}) \models \bigwedge_{i<j\leq N} \Phi_j 
		\end{array} 
		\right\}.\nonumber
	\end{align}

	Also, we define $\mathcal{T}_{N}=\mathcal{X}$. 
	Later in Section~\ref{sec:feasibleset}, we will discuss in detail how to compute terminal sets $\mathcal{T}_i$. 
	Now, in the context of shrinking horizon MPC, 
	set $\mathcal{T}_i$ can be considered as the \emph{terminal constraint} in the optimization problem for sub-task $\Phi_i$. 
	Therefore, \texttt{Problem~1} is further modified as the following optimization problem with terminal constraint. 
	
	\begin{myprob}\label{prob2}\upshape
		\textbf{(Robust STL Optimization Problem with Terminal Constraint).}
		Given system in the form of  Equation~(\ref{eq:system}), an STL formula $\Phi$, a cost function $J$, 
		the current state $x_k \!\in\! \mathcal{X}$ at instant $k$, 
		time horizon $[S,T]$,
		previously state sequence $\mathbf{x}_{S:k-1}$, 
		and a terminal set $\mathcal{T}$, 
		find an optimal input sequence $\mathbf{u}^*_{k:T-1}$ that minimizes the cost function subject to  	constraints on the system's dynamic, the temporal logic requirement and the terminal constraint. 
		Formally, we have the following optimization problem
		\begin{subequations}\label{second:main2} 
			\begin{align}
				& \underset{\mathbf{u}_{k:T-1}}{\text{minimize}}  & & J(x_k, \mathbf{u}_{k:T-1} )\\
				& \text{subject to}\nonumber
				& & \\
				& &&
				\!\!\!\!\!\!\!\!\!\!\!\!\!\!\!\!\!\!\!\!\!\!\!\!\!\!\!\!\!\!\!
				\forall \mathbf{w}_{k:T-1}\!\in\! \mathcal{W}^{T \shortminus k}: \mathbf{x}_{S:k} \xi_f(x_k,\mathbf{u}_{k:T \shortminus 1},\mathbf{w}_{k:T \shortminus 1})\models \Phi, \\
				& &&
				\!\!\!\!\!\!\!\!\!\!\!\!\!\!\!\!\!\!\!\!\!\!\!\!\!\!\!\!\!\!\!
				\forall \mathbf{w}_{k:T-1}\!\in\! \mathcal{W}^{T \shortminus k}: 
				\xi_{f}^T(x_k,\mathbf{u}_{k:T \shortminus 1},\mathbf{w}_{k:T \shortminus 1}) \in \mathcal{T}, \\
				& &&
				\!\!\!\!\!\!\!\!\!\!\!\!\!\!\!\!\!\!\!\!\!\!\!\!\!\!\!\!\!\!\!\!
				u_k,u_{k+1},\cdots, u_{T-1}\in \mathcal{U}.
			\end{align} 
		\end{subequations} 
	\end{myprob} 
	\begin{remark}\upshape 
	Similar to  Problem~\ref{prob1}, the CEGIS-based approach can be also applied to solve Problem~\ref{prob2} with the terminal constraint. Basically, CEGIS uses inductive counterexamples of the disturbance sequence to deal with \emph{forall} constraint to compute the optimal solution. Therefore, the  terminal constraint (\ref{second:main2}c) added does affect the solution process. In our algorithm implementation, we still adopt the CEGIS-based method. 
	\end{remark}
	\subsection{Overall Framework}
	
	\begin{algorithm}[ht]
		\caption{Shrinking Horizon MPC for STL with Time Interval Decomposition}
		\label{alg:cap}
		\KwIn{STL formula $\Phi$ of form Equation~\eqref{eq:separation},  dynamic system model of form ($\ref{eq:system}$) and   cost function $J$ }
		\KwOut{control input $u_k$ at each instant $k$.\vspace{3pt}}
		for each $i=1,\cdots,N-1$, compute $\mathcal{T}_i$ \\
		$i \gets 1$, $ k \gets 0$ and $T_0=0$\\
		\While{$i \leq N$}
		{
			$\Phi \gets \Phi_i, S \gets T_{i-1}, T \gets T_i$

			\While{$k < T$}
			{
				measure current state $x_k$ \\
				solve \texttt{Problem~2} based on $\mathbf{x}_{S:k}$, $T$,  $\Phi$ and $\mathcal{T}_i$ and  obtain optimal inputs $\mathbf{u}_{k:T\shortminus 1}^*= u_k^*u_{k+1}^*  \cdots u_{T \shortminus 1}^*$ \\
				apply control input $u_k^*$ \\
				$k \gets k+1$ 
			}
			$i \gets i+1$
		}
	\end{algorithm}

	The overall time interval decomposition framework is formalized by \texttt{Algorithm~2}.
	Specifically, line 1 constructs the terminal constraints in an offline manner for each online optimization stage. The computation  will be detailed in the next section. 
	Lines 2-10 aim at synthesizing control strategy for each sub-formula iteratively. 
	For each specific sub-formula (lines 5-9), we use the shrinking horizon model predictive control and apply the first element of the control input sequence to the system which is similar to \texttt{Algorithm~1}.  Note that line 1 is executed offline and the remaining procedures are computed online to resist uncertain disturbances.
	
	The following result shows that the completeness and correctness of   \texttt{Algorithm~2}.
	\begin{thm}\upshape
		Given dynamic system model (\ref{eq:system}), STL formula $\Phi$, cost function $J$ and initial state $x_0$, if we can obtain a control input sequence $\mathbf{u}^*_{0:T_1-1}$ at instant $k=0$ by \texttt{Algorithm~2}, then \texttt{Algorithm~2} is feasible for all the time instants and the solution with returned $\mathbf{u}^* = [u_0^*, \cdots, u_{T_\Phi-1}^*]$ satisfies the STL formula $\Phi$.
	\end{thm}
	\begin{pf}
		The existence of an initial control input sequence $\mathbf{u}^*_{0:T_1-1} = u_{0}^* u_{1}^* \cdots u_{T_1-1}^*$ implies that for all possible disturbance sequences $\mathbf{w}_{0:T_1-1}$, the optimization problem is feasible. 
		At the next time step $k=1$, there  exists at least one sub-optimal input sequence $\mathbf{u}_{1:T_1-1} = u_{1|0}^* \cdots u_{T_1-1|0}^*$ obtained last time step that can resist all possible disturbances, i.e., the robust problem at instant $k=1$ is feasible. Then we can prove it recursively in the first stage.
		
		For the second stage, we have that, from  state $x_{T_1}$, there exists a control sequence, e.g., $\bar{\mathbf{u}}_{T_1:T_{\Phi}-1} \in \mathcal{U}^{T_{\Phi}-T_1}$, such that $\mathbf{x}_{T_1+1:T_{\Phi}} \models \bigwedge_{1<i\leq N} \Phi_i$ by the definition of terminal constraint (\ref{eq:terminal}) since state $x_{T_1}$ is in the terminal set $\mathcal{T}_{1}$ by constraint (\ref{second:main2}c) at the first stage.	
		As a result, at time $T_1$, the first time in the second stage, when solving the optimization problem (\ref{second:main2}), we can at least find a feasible control input sequence $\bar{\mathbf{u}}^*_{T_1:T_2-1}$ to satisfy (\ref{second:main2}b), (\ref{second:main2}c) and (\ref{second:main2}d).
		Then, the same as stage 1, the problem is feasible in the second stage. Analogously, we can prove the recursive feasibility in the remaining stages.
		
		Furthermore, at the last time step of each stage, constraint (\ref{second:main2}b) ensures that the state sequence $\mathbf{x}_{T_{i-1}+1:T_i}$ of the current stage $i$ meets the requirements of sub-formula $\Phi_i, i \in \{ 1,\cdots, N \}$. Therefore, STL formula $\Phi = \Phi_1 \wedge \cdots \wedge \Phi_N$ is also satisfied.
		\hfill		$\qed$
	\end{pf}
	
	\begin{remark}
	    Let us discuss the computational advantage for the proposed algorithm. 
	    First of all, the complexity for computing terminal constraints in (\ref{second:main2}c) largely depends on the system model and STL formula. Nevertheless, this computation is preformed fully offline and will not be the bottleneck of the online computations. 
	    Regarding the online computation, in general, solving \texttt{Problem~2} is exponential in the length of the prediction horizon. Here, since the time horizon is decomposed, the complexity for  online optimization is significantly reduced. 
		The finer the formula can be decomposed, the more complex reduction one can obtain. 
	\end{remark}
	
	\begin{remark}
		Finally, we remark that, if terminal sets $\mathcal{T}_{i}, i\!\in\! \{1, \cdots, N\}$ can be computed precisely, then our framework is both sound and as complete as classic MILP method with the consideration of disturbance. 
		In practice, one may need to compute the inner-approximations of terminal sets. 
		In this case, our framework is sound but not complete since the terminal sets are smaller than what are needed.
	\end{remark}

\section{Computation of Terminal Constraints} \label{sec:feasibleset}

In this section, we provide details for the computations of  terminal sets as defined in Equation~\eqref{eq:terminal},  which are essentially feasible set of subsequent tasks taking disturbances into account. 
A direct approach to compute this set is to use a branch-and-bound algorithm to search for the terminal set as the framework in \cite{bravo2005computation}. However, using this method, at each branch, we need to determine if \emph{there exists} control input sequence  $\mathbf{u}_{T_i:T_\Phi-1}$ such that \emph{for all} disturbance  sequence  $\mathbf{w}_{T_i:T_\Phi-1}$  the subsequent formula can be satisfied, which is very computational challenging.

Here, instead of computing $\mathcal{T}_i$ explicitly, we seek to compute its inner-approximation $\hat{\mathcal{T}}_i\subseteq \mathcal{T}_i$. In terms of the MPC problem, the satisfaction of 
$\hat{\mathcal{T}}_i$ implies the satisfaction of $\mathcal{T}_i$. Although considering the inner-approximations of the terminal sets is a bit more conservative, the computation can be done much more efficiently.

Specifically, we define the following set $\hat{\mathcal{T}}_{i}$  to approximate terminal set $\mathcal{T}_i$
 \begin{align}\label{eq:terminalequi}
		&\hat{\mathcal{T}}_{i} =  \\
		&\left\{
		x_{T_{i}} \!\in\! \mathcal{X} \,\middle\vert\,
		\begin{array}{cc}
			\exists u_{T_i} \forall w_{T_i} \!\in\! \mathcal{W} \ \exists u_{T_i+1} \forall w_{T_i+1} \!\in\! \bigoplus_{i=0}^{1}L^i \mathcal{W}\\
			\cdots \exists u_{T_{\Phi} \shortminus 1}  \forall w_{T_{\Phi} \shortminus 1} \!\in\!  \bigoplus_{i=0}^{T_\Phi-T_i-1}L^i \mathcal{W}  \\
			\text{s.t. }  	\xi_f(x_{T_{i}}, \mathbf{u}_{T_{i}:T_{\Phi} \shortminus 1}, \mathbf{w}_{T_{i}:T_{\Phi}\shortminus 1}) \models \bigwedge_{i<j\leq N} \Phi_j 
		\end{array} 
		\right\}. \nonumber
	\end{align}
The above defined  set $\hat{\mathcal{T}}_{i}$ differs from the original terminal set $\mathcal{T}_i$ in two folds: 
\begin{itemize}[leftmargin=*] 
    \item 
    First, in the definition of $\mathcal{T}_i$, we require the existence of an open-loop control input sequence robust to all possible disturbance sequences. 
    However, in the definition of $\hat{\mathcal{T}}_i$, the control inputs can be determined after observing each specific states (or disturbances). 
    Therefore, there are alternations between the existential quantifiers and the universal quantifiers.
    This allows us to compute the set inductively in a backward manner. \medskip
    \item  
    Second, the disturbance set grows at each instant according to the  Lipschitz constant. 
    Specifically, set $\bigoplus_{i=0}^{k}L^i \mathcal{W} $  upper bounds  possible disturbances of the system at instant $T_i+k$.  
    Clearly, this set is easy to compute but, in general, larger than the actual disturbance to the system.  
    Therefore, the corresponding terminal set is smaller than the actual one. 
\end{itemize}
 

The following result formally shows that 	$\hat{\mathcal{T}}_{i} $ is indeed an inner-approximation of    $\mathcal{T}_{i}$.

\begin{mypro}\label{lem:terminalequi}
For each terminal set $\mathcal{T}_i$,   we have $\hat{\mathcal{T}}_{i}  \subseteq \mathcal{T}_i$.
\end{mypro}
\begin{pf}
The proof can be found in Appendix \ref{app:pro}.
\end{pf}

Then, we present the computation methods of $\hat{\mathcal{T}}_i$ for each $i \!\in\! \{1, \cdots, N-1\}$ as follows with the help of $I$-remaining \emph{robust} feasible sets whose notion is similar to $I$-remaining feasible sets in Definition \ref{def:feasibleset}.
 
\begin{mydef}[$I$-Remaining Robust Feasible Sets]\label{def:robustset}\upshape
	Given an STL formula $\Phi$ of form~\eqref{eq:stl-seg}, 
	a subset of indices $I\subseteq \mathcal{I}$, a starting instant $s$ and current instant $k$ with $s \leq k$, 
	then starting from instant $s$, $I$-remaining robust feasible set at $k$, denoted by 
	$\hat{X}_{s,k}^{I} \subseteq \mathcal{X}$, is the set of states as follows,
	\begin{align}\label{eq:feasibleset}
		& \hat{X}_{s,k}^{I} =  \\
		& \left\{ 
		x_k \!\in\! \mathcal{X} \,\middle\vert\, \!\!\!\!
		\begin{array}{cc}
			\exists u_k \forall w_k \!\in\! \bigoplus_{i=0}^{k\shortminus s} L^i\mathcal{W} \ 
			\cdots \\
			\ \exists u_{T_{\Phi} \shortminus 1}  \forall w_{T_{\Phi} \shortminus 1} \!\in\!  \bigoplus_{i=0}^{T_\Phi \shortminus 1 \shortminus s} L^i\mathcal{W} \\
			\text{ s.t. } x_k\xi_f(x_{k}, \mathbf{u}_{k:T_\Phi-1},  \mathbf{w}_{k:T_{\Phi}\shortminus 1}) \models \hat{\Phi}_{k}^I
		\end{array} 
		\right\}. \nonumber
		\vspace{-6pt}
	\end{align}
\end{mydef}

In this above definition,  parameter $s$ represents the time instant when a sub-formula starts. 
This information is used in order to determine how many times the disturbance set should be magnified. 
Then parameter $k$ is still the current instant from which the control sequence is applied.  
Then by definition, when $s=k=T_i$, the $I$-remaining robust feasible set is  indeed the inner-approximation of terminal set that we want, i.e., 
\begin{equation}
    \hat{\mathcal{T}}_i = \hat{X}_{T_i, T_i}^{\{i+1, \cdots, N\}}. \nonumber
\end{equation}

For the computation of $\hat{X}_{s,k}^{I}$, we can apply a similar approach for computing $X_k^I$ in Equation~\eqref{eq:dynpro}. 
Specifically, for each fixed $s$, we use $\hat{X}_{s,k+1}^{I'}$ to compute $\hat{X}_{s,k}^{I}$ in a backwards manner until $k=s$. 
The computation is summarized by the following theorem. 

\begin{thm}\label{thm:terminal}
For each time instant $s$, 
the	$I$-remaining robust feasible set $\hat{X}_{s,k}^{I}$  defined in Definition~\ref{def:robustset} for  time instant $k$ can be computed as follows
	\begin{equation}\label{eq:rX}
		\hat{X}_{s,k}^{I} = \bigcup_{I' \in \textsf{succ}(I, k)} \Big( H_{k}(I, I') \cap \Upsilon_r(\hat{X}_{s,k+1}^{I'}, h(\mathcal{W}, s)) \Big), 
	\end{equation} 
	where $h(\mathcal{W}, s ) = \bigoplus_{i=0}^{k-s} L^i \mathcal{W}$ and $\Upsilon_r(\cdot, \cdot)$ is the robust one-step set defined by: for any $\mathcal{S}\subseteq \mathcal{X}$, we have
	\begin{equation}\label{eq:one-step-f}
		\Upsilon_r(\mathcal{S}, \mathbf{W}) = \{x \in \mathcal{X} \mid \exists u \in \mathcal{U} \ \forall w \in \mathbf{W}  \text{ s.t. }  f(x, u) + w \in \mathcal{S}\}. \nonumber
	\end{equation}
\end{thm}

\begin{pf}
The proof can be found in Appendix \ref{app:thm}.
\end{pf}

\section{Case Study}\label{sec:case}
In this section, we apply the proposed time interval decomposition framework to a case study of planar motion of a single robot with double integrator dynamics.
Online  simulations are conducted in \textsf{Python} 3 and we use \textsf{Gurobi} \cite{optimization2018gurobi} to solve the optimization problem. 
The terminal constraints are computed offline in \textsf{Julia} with the help of existing packages \textsf{JuliaReach} \cite{lazysets21, bogomolov2019juliareach}.
All simulations are carried out by a laptop computer with i7-10510U CPU and 16 GB of RAM.
Our codes are available at \url{https://github.com/Xinyi-Yu/MPC4STL-TID}, where more details can be found.

\textbf{System Model}:
The model with a sampling period of 0.5 seconds is as follows,
\[
    x_{k+1} = A x_k + B u_k + w_k,
\]
where 
\[
A = \begin{bmatrix}
		1 & 0.5 & 0 & 0 \\
		0 & 1 &  0 & 0 \\
		0 & 0 & 1 & 0.5 \\
		0 & 0 & 0 & 1 
	\end{bmatrix}
	\text{ and }B = \begin{bmatrix} 
		0.125 & 0   \\ 
		0.5 & 0 \\ 
		0 & 0.125 \\ 
		0 & 0.5 
	\end{bmatrix}
\]
and state $x_k = [x \ v_x \ y \ v_y]^T$ denotes $x$-position, $x$-velocity, $y$-position and $y$-velocity, and control input $u_k = [u_x \ u_y]^T$ denotes $x$-acceleration and $y$-acceleration respectively. The physical constraints are 
\[
x \!\in\! \mathcal{X} = [0,10] \times [-2.5, 2.5] \times [0,10] \times [-2.5, 2.5],
\]
and $u \!\in\! \mathcal{U} = [-3, 3]^2$.
The disturbance $w_k = [w_x \ w_{vx} \ w_y \ w_{vy}]^T$ is assumed to be bounded by a compact set $[-0.01, 0.01]^4$.
 
\begin{figure}
	\centering
	{\includegraphics[width=8cm]{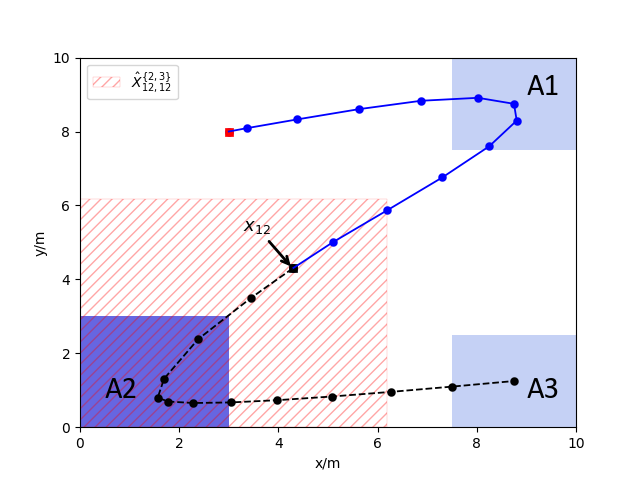}}
	\caption{The result of simulation trace.}
	\label{fig:Simulation}
\end{figure}

\textbf{Planning Objectives}: 
We assume that the initial state of the robot is $x_0 = [3, 0, 8, 0]$ shown as the red point in Fig.~\ref{fig:Simulation}.
The control objective of the robot is to visit region $A_1$ at least once between instants 0 to 6 (from 0s to 3s), always stay at region $A_2$ between instants 14 to 15 (from 7s to 7.5s) and finally reach region $A_3$ at least once between instants 22 to 25 (from 11s to 12.5s). Such an objective can be specified by  STL formula as follows:
\[
\Phi = \textbf{F}_{[0,6]}A_1 \wedge \textbf{G}_{[14,15]} A_2 \wedge \textbf{F}_{[22, 25]} A_3,
\]
where $A_1 = (x \in [7.5, 10]) \wedge (y \in [7.5, 10])$, $A_2 = (x \in [0,3]) \wedge (y \in [0,3])$ and $A_3 = (x \in [7.5, 10]) \wedge (y \in [0, 2.5])$.

\textbf{Simulation Results}: 
In our   framework, we can decompose the formula $\Phi$ into $\Phi_1^{[0,12]}$ and $\Phi_2^{[13,25]}$ as follows: 
\begin{equation}
	\begin{array}{ll}
		\Phi_1^{[0,12]} = \textbf{F}_{[0, 6]}A_1 , \\
		\Phi_2^{[13,25]} = \textbf{G}_{[14,15]} A_2 \wedge \textbf{F}_{[22, 25]} A_3. \nonumber
	\end{array}
\end{equation}
For this case, there is only one terminal set at instant 12 needs to be computed, whose inner approximation  is shown as the red shaded part in Fig.~\ref{fig:Simulation}. 
The online executed trajectory is also shown in Fig.~\ref{fig:Simulation}. 
Specifically, the simulation result of the  first stage  is represented by blue dots and solid line; 
the simulation result of the  second stage is printed by black dots and dashed line. 
Clearly, the black rectangle dot  $x_{12}$ falls into the  approximated terminal set $\hat{X}_{12,12}^{\{2,3\}}$ in order to ensure the continuation of the task.
Note that the offline results shown in the figure is the projection to the first and third dimensions but the terminal set is still constrained in the complete 4-dimensional state space. 

In terms of computation time and performance, 
we compared  our method with the standard MILP method in \cite{raman2014model}. 
The comparison result is shown in Table 1, where the cost function we used  is 
\begin{align}
    & J(x_k, \mathbf{u}_{k:T-1}) = \nonumber \\  
	& -\rho^\Phi_{  \mathbf{x}_{0:k} \xi_f(x_k,\mathbf{u}_{k:T-1}),k} + 0.67 \times 10^{-7} \times \Sigma_{i=k}^{T-1} (u_x^2 + u_y^2). \nonumber
\end{align}
Note that all results in Table 1 are  averaged results among 20 simulations.\footnote{Here, we use CEGIS to find a control input sequence which can resist all possible disturbances. The  efficiency of this method relies on the given initial disturbance sequences. Therefore, we use average time of multiple simulations to better justify the computation time and performance.} 
Compared with the  standard MILP approach, one can see that the cost of the trajectory using our approach is slightly higher. 
This is because we do not solve the optimization problem globally. 
However, using our approach, the computation time for the online optimization problem is significantly smaller than that of the MILP approach, especially for the initial stages.   In particular, the  MILP approach cannot keep pace with the sampling time $0.5$s of the system. However, our approach can ensure online control for this system.

\begin{table}[htb] 
\caption{Comparison of simulation results}  
\label{table:1} 
\begin{tabular}{|c|c|c|c|c|c|}
\hline   
\multirow{2}{*}{\textbf{Algorithms}} & \multicolumn{4}{c|}{\textbf{Computation time}}  & \multirow{2}{*}{\textbf{Cost}}
 \\
\cline{2-5}
 & k=0 (s) & k=1 (s)  & ... & total (s) &  \\
\hline
\makecell{\textbf{Time-Interval} \\ \textbf{Decomposition}} &0.1989&0.0488 & ... &1.4367 &-1.2299 \\
\hline
\makecell{\textbf{General MILP} \\ \textbf{Framework}} & 41.28& 6.58& ... &82.83 & -1.2478\\
\hline
\end{tabular}   
\end{table}

\section{Conclusion}\label{sec:conclu}
In this work,  we proposed a new framework for model predictive control of STL specifications. 
We showed that, by effectively computing the terminal set of each subsequent sub-formula, the long horizon MPC problem can be decomposed into a set of MPC problems with shorter horizon. The computational efficiency of the proposed framework was illustrated by a case study of mobile robot.  
In this work, we assume that the STL formula can be naturally decomposed into several sub-formulae with disjoint time intervals. 
In the future, we aim to relax this assumption by tackling the case with any receding prediction horizon.  

\bibliography{STL} 

\newpage
\appendix

\section{Proof of Proposition 1}\label{app:pro}
We prove the proposition by showing that if $x_{T_i} \!\in\! \hat{\mathcal{T}}_{i}$, then $x_{T_i} \!\in\! \mathcal{T}_{i}$ as follows.

\textbf{Step 1:} Given that $x_{T_i} \!\in\! \hat{\mathcal{T}}_{i}$, assume that the existing control inputs are 
\begin{equation}\label{eq:hatu}
    \begin{split}
        \hat{u}_{T_i} &= h_0(x_{T_i}),\\
        \hat{u}_{T_i+1} &= h_1(x_{T_i}, w_{T_i}),\\
        &\cdots,\\
        \hat{u}_{T_\Phi-1} &= h_{T_\Phi - T_i - 1}(x_{T_i}, w_{T_i}, \cdots, w_{T_\Phi-2}),
    \end{split}
\end{equation}
where $\{h_0, h_1, \cdots, h_{T_\Phi - T_i - 1}\}$ is a kind of control strategy for $x_{T_i}$ and $w_{T_i+p} \!\in\! \bigoplus_{i=0}^{p}L^i \mathcal{W}, p=0, \cdots, T_\Phi - T_i - 2$.
Recall that in $\hat{\mathcal{T}}_i$ case, the input $\hat{u}_{T_i+p}$ is decided after observing previous information which is captured by the elements in function ``$h$". 
There could be more than one strategy $\{h_0, h_1, \cdots, h_{T_\Phi - T_i - 1}\}$, leading to more than one control input sequence $\hat{\mathbf{u}}_{T_i:T_\Phi-1}$, and we denote by $\mathcal{H}$ the set of all the possible control strategy sequence for state $x_{T_i}$. 

\textbf{Step 2:} In set $\mathcal{H}$, there definitely exists at least one control strategy $\{h_0^*, h_1^*, \cdots, h^*_{T_\Phi - T_i - 1}\}$ in which the specific strategy of every step will attenuate all the previous disturbance, i.e., for all $p \in \{1, \cdots, T_\Phi-T_i-2\}$ and for all $w_{T_i+j} \!\in\! \bigoplus_{i=0}^{j}L^i \mathcal{W}, j=0, \cdots, p-1$ we have 
\begin{align}\label{eq:property}
    & f(x_{T_i+p},h_p^*(x_{T_i},\mathbf{0}^n,\cdots,\mathbf{0}^n) \in \nonumber  \\
    & f(x_{T_i+p}, h^*_p(x_{T_i},w_{T_i},\cdots,w_{T_i+p-1}))  \oplus \bigoplus_{i=1}^p L^i\mathcal{W},
\end{align}
where $x_{T_i+p}$ is a real state with disturbance at instant $T_i+p$ and $\bigoplus_{i=1}^p L^i\mathcal{W}$ is all the possible influences to $x_{T_i+p+1}$ caused by disturbances of first $p$ steps. 
The reason is as follows. (Of course, the attenuating thing is also what we want function $h$ to do.) Suppose that there exists a control strategy $\{h'_0, h'_1, \cdots, h'_{T_\Phi - T_i - 1}\} \in \mathcal{H}$ and in this strategy some functions $h'_q, q \in Q \subseteq \{1, \cdots, T_\Phi - T_i - 2\}$ does not attenuate previous disturbance. It implies that under these function $h'_q, q \in Q$, there exists a disturbance sequence $\{w'_{T_i},\cdots,w'_{T_i+q-1}\}$ such that 
\begin{align}
    & f(x_{T_i+q},h_q'(x_{T_i}, \mathbf{0}^n,\cdots,\mathbf{0}^n)  \notin \nonumber \\
    & f(x_{T_i+q}, h'_q(x_{T_i},w'_{T_i},\cdots,w'_{T_i+q-1})) \oplus \bigoplus_{i=1}^q L^i\mathcal{W}. \nonumber
\end{align}
Under this condition, there definitely exists such a control strategy $\{h_0^*, h_1^*, \cdots, h^*_{T_\Phi - T_i - 1}\}$ which will attenuate all the previous disturbances, where 
\begin{equation}
    h_j^* = \left\{
    \begin{aligned}
        & h_j'(x_{T_i},w_{T_i},\cdots,w_{T_i+j-1})  & \text{if} \ j \notin Q \\
        & h_j'(x_{T_i},\mathbf{0}^n,\cdots,\mathbf{0}^n).  & \text{if} \ j \in Q \\
    \end{aligned}
    \right. \nonumber
\end{equation}

\textbf{Step 3:} For such a state $x_{T_i}$, we claim that $x_{T_i}\in \mathcal{T}_i$ since there exists at least a control strategy $\{h_0^*, h_1^*, \cdots, h^*_{T_\Phi - T_i - 1}\}$ such that for all disturbance sequences the solution is right, i.e., the control inputs are
\begin{equation} \label{eq:u}
    \begin{split}
        u_{T_i} &= h^*_0(x_{T_i}),\\
        u_{T_i+1} &= h^*_1(x_{T_i}, \mathbf{0}^n),\\
        &\cdots,\\
        u_{T_\Phi-1} &= h^*_{T_\Phi - T_i - 1}(x_{T_i}, \mathbf{0}^n, \cdots, \mathbf{0}^n).
    \end{split} 
\end{equation}
The proof is as follows. 
We first clarify two different control modes corresponding to $\mathcal{T}_i$ and $\hat{\mathcal{T}}_i$ respectively as follows: \vspace{-6pt}
\begin{itemize}
    \item fixed control sequence $u_{T_i}, u_{T_i+1}, \cdots, u_{T_\Phi-1}$ as \eqref{eq:u} with random disturbance sequence $\forall \mathbf{w}_{T_{i}:T_{\Phi}\shortminus 1} \!\in\! \mathcal{W}^{T_{\Phi} \shortminus T_{i}} $ and
    \item $\hat{u}_{T_i}, \forall w_{T_i} \!\in\! \mathcal{W}, \hat{u}_{T_i+1}, \forall w_{T_i+1} \!\in\! \bigoplus_{i=0}^{1}L^i \mathcal{W} \ \cdots  \  \hat{u}_{T_{\Phi} \shortminus 1}, \\
    \forall w_{T_{\Phi} \shortminus 1} \!\in\!  \bigoplus_{i=0}^{T_\Phi-T_i-1}L^i \mathcal{W}$ where $\hat{u}_{T_i}, \hat{u}_{T_i+1}, \cdots, \hat{u}_{T_\Phi-1}$ is defined in \eqref{eq:hatu}.
\end{itemize}
  
We introduce two symbols $\{x_{T_i+p+1}\}_p$ and $\{\hat{x}_{T_i+p+1}\}_p$ which represent the reachable sets starting from the same state $x_{T_i+p}$ by applying above two strategy respectively with consideration of disturbance at instant $T_i+p$, i.e., 
\begin{equation}
    \left\{
    \begin{aligned}
        \{x_{T_i+p+1}\}_p = & f(x_{T_i+p}, h_p(x_{T_i}, \mathbf{0}^n,\cdots,\mathbf{0}^n))\oplus \mathcal{W}\\
        \{\hat{x}_{T_i+p+1}\}_p = & f(x_{T_i+p}, h_p(x_{T_i},w_{T_i},\cdots,w_{T_i+p})) \\
       & \oplus L^{p}\mathcal{W} \oplus \cdots\oplus \mathcal{W}.\\
    \end{aligned}
    \right. \nonumber
\end{equation}
According to Equation \eqref{eq:property}, we have $\{x_{T_i+p+1}\}_p \subseteq \{\hat{x}_{T_i+p+1}\}_p$ obviously, which implies that with consideration of disturbance, starting from the same state, the reachable set after applying the former control way will be contained in that after applying the latter. 
It is easy to induce that the set of all the possible state sequence starting from $x_{T_i}$ under the former control mode, denoted by $\{\mathbf{x}_{T_i:T_\Phi}\}$, is contained in that under the latter, denoted by $\{\hat{\mathbf{x}}_{T_i:T_\Phi}\}$, i.e., $\{\mathbf{x}_{T_i:T_\Phi}\} \subseteq \{\hat{\mathbf{x}}_{T_i:T_\Phi}\}$. 

Since $x_{T_i} \!\in\! \hat{\mathcal{T}}_{i}$, we have $\forall \hat{\mathbf{x}}_{T_i:T_\Phi} \in \{\hat{\mathbf{x}}_{T_i:T_\Phi}\}$, $\hat{\mathbf{x}}_{T_i:T_\Phi} \models \bigwedge_{i<j\leq N} \Phi_j$. Then we can obtain $\forall \mathbf{x}_{T_i:T_\Phi} \in \{\mathbf{x}_{T_i:T_\Phi}\}, \mathbf{x}_{T_i:T_\Phi} \models \bigwedge_{i<j\leq N} \Phi_j$. Therefore, $x_{T_i}\in \mathcal{T}_i$ holds, i.e., the proposition is proved.

\section{Proof of Theorem 3}\label{app:thm}
\begin{figure*} \tiny
	\begin{align} \label{eq:proof-equ}
		\hat{\Phi}_{k}^{I} = \bigvee_{\hat{I} \subseteq I_k^\mathbf{U} \cap \mathcal{I}_{k+1}} 
		\Big(
			\underbrace{
			\bigwedge_{i \in I_k \setminus \mathcal{I}_{k+1}} \Phi_i^{[k,k]} 
			\wedge \bigwedge_{i \in I_k^\mathbf{U} \cap \mathcal{I}_{k+1} \setminus \hat{I}} \Phi_i^{[k,k]}
			\wedge \bigwedge_{i \in \hat{I}} \mathbf{G}_{[k,k]}x \!\in\! \mathcal{H}_i^1
			\wedge \bigwedge_{i \in I_k^\mathbf{G} \cap \mathcal{I}_{k+1}} (\Phi_i^{[k,k]} 
			}_{\psi_1(\hat{I})}
			\wedge
			\underbrace{
			\Phi_i^{[k+1,b_i]})
			\wedge \bigwedge_{i \in \hat{I}} \Phi_i^{[k+1,b_i]}
			\wedge \bigwedge_{i \in \mathcal{I}_{>k}} \Phi_i^{[a_i,b_i]}
			}_{\psi_2(\hat{I})}
		\Big)
	\end{align}
\end{figure*}

    When $k = T$, since $\textsf{succ}(I, k) = \{\emptyset\}$ and $\textsf{sat}_\textsf{U}(I,I') = \textsf{sat}_\textsf{U}(I,\emptyset)  = \{i \!\in\! I: O_i = \mathbf{U}'\}$, from Equation~\eqref{eq:rX}, we have
    \[
    \hat{X}_{s,T}^I = \bigcap_{i \in \textsf{sat}_\textsf{U}(I,I')} (\mathcal{H}_i^1 \cap \mathcal{H}_i^2) \cap \bigcap_{i \in I \setminus \textsf{sat}_\textsf{U}(I,I')} \mathcal{H}_i,
    \]
    which is clearly the $I$-remaining robust feasible set of $\hat{\Phi}_{k}^{I} = \bigwedge_{i \in I} \Phi_i^{[T,T]}$.

	For the case of $k \neq T$, we can always write 
	$\hat{\Phi}_{k}^{I}$ as Equation \eqref{eq:proof-equ}, where $I_k$ is the abbreviation of $I\cap \mathcal{I}_k$, and $I_k^\mathbf{G}, I_k^\mathbf{U}$ are the sets of ``Always'' and ``Until'' indices contained in $I_k$, respectively.
	Intuitively, Equation~\eqref{eq:proof-equ}  divides the corresponding formula into two parts: 
	one only related to the current state (denoted by $\psi_1(\hat{I})$) and the other related to the future requirements (denoted by $\psi_2(\hat{I})$), where $\hat{I}$ is the index set for those ``Until" sub-formulae that are not satisfied currently, which can be any subset of $I_k^\mathbf{U}\cap \mathcal{I}_{k+1}$. The detailed deduction can be found in Equation (15) in \cite{yu2022model}. 
	
	We observe that, for any  $\hat{I}$, we have 
	\begin{align}
	& \exists u_k \forall w_k \!\in\! \bigoplus_{i=0}^{k\shortminus s} L^i\mathcal{W} \ 
	\cdots \ \exists u_{T_{\Phi} \shortminus 1}  \forall w_{T_{\Phi} \shortminus 1} \!\in\!  \bigoplus_{i=0}^{T_\Phi \shortminus 1 \shortminus s} L^i\mathcal{W}, \nonumber \\
	& \text{s.t.} \ x_k \xi_f(x_{k}, \mathbf{u}_{k:T_\Phi-1},  \mathbf{w}_{k:T_{\Phi}\shortminus 1}) \models \psi_1(\hat{I}) \wedge \psi_2(\hat{I}),
	\nonumber
	\end{align}
	if the following holds: \vspace{-6pt}
	\begin{itemize}
		\item $x_k \models \psi_1(\hat{I})$, \vspace{-6pt}
		\medskip
		\item $\exists u_{k+1} \forall w_{k+1} \!\in\! \bigoplus_{i=0}^{k\shortminus s+1} L^i\mathcal{W} \ 
		\cdots \ \exists u_{T_{\Phi} \shortminus 1}  \forall w_{T_{\Phi} \shortminus 1} \!\in\!  \bigoplus_{i=0}^{T_\Phi \shortminus 1 \shortminus s} L^i\mathcal{W}, \ \text{s.t.} \ x_{k+1} \xi_f(x_{k+1}, \mathbf{u}_{k+1:T_\Phi-1},  \mathbf{w}_{k+1:T_{\Phi}\shortminus 1}) \\
		\models \psi_2(\hat{I})$, \vspace{-6pt}
		\medskip
		\item $x_{k+1}$ is reachable from $x_k$ under some $u_k$ regardless of any disturbances $w_k$.
	\end{itemize}
	The first condition holds iff  $x_k$ stays in region $H_k(I, I')$ with $I' = \mathcal{I}_{>k} \cup (I_k^\mathbf{G} \cap \mathcal{I}_{k+1}) \cup \hat{I} $ and $\textsf{sat}_\textsf{U}(I, I') = (I_k^\mathbf{U} \setminus \mathcal{I}_{k+1}) \cup (I_k^\mathbf{U} \cap \mathcal{I}_{k+1} \setminus \hat{I})$. 
	The second condition holds iff  $x_{k+1}$ is in $I'$-remaining feasible set $\hat{X}_{s,k+1}^{I'}$. 
	The third condition holds iff  $x_k \!\in\! \Upsilon_r(\hat{X}_{s,k+1}^{I'}, \bigoplus_{i=0}^{k-s} L^i \mathcal{W})$.
	Therefore,  $\psi(\hat{I}):=\psi_1(\hat{I}) \wedge \psi_2(\hat{I})$ holds iff  $x_k$ is in $H_{k}(I, I') \cap \Upsilon_r(\hat{X}_{s,k+1}^{I'}, h(\mathcal{W}, s))$ with $h(\mathcal{W}, s ) = \bigoplus_{i=0}^{k-s} L^i \mathcal{W}$. 	
	Finally, recall that $\hat{\Phi}_{k}^{I}$ is the disjunction of all possible $\psi(\hat{I})$ and this suffices to consider all possible $I' \in \textsf{succ}(I,k)$.
	Therefore, we have $\hat{X}_{s,k}^{I} = \bigcup_{I' \in \textsf{succ}(I, k)} \Big( H_{k}(I, I') \cap \Upsilon_r(\hat{X}_{s,k+1}^{I'}, h(\mathcal{W}, s)) \Big)$ which is the same as Equation~\eqref{eq:rX}, i.e., the theorem is proved.

\end{document}